\documentclass[reprint,superscriptaddress,amsmath,amssymb,aps,]{revtex4-2}
\usepackage{graphicx}
\usepackage{dcolumn}
\usepackage{bm}
\usepackage{hyperref}
\usepackage{amsmath}
\usepackage{tabularx}
\usepackage{graphicx}
\usepackage{dcolumn}
\usepackage{bm}
\usepackage{cleveref}
\usepackage{xcolor}

\begin{document}
\preprint{APS/123-QED}
\title{Elasto-inertial force exerted on a sphere oscillating in a fluid near a soft interface}
\author{Axel Batalha}
\affiliation{Univ. Bordeaux, CNRS, LOMA, UMR 5798, 33405 Talence, France.}
\affiliation{Laboratoire Charles Coulomb (L2C), Univ. Montpellier, CNRS, Montpellier, France.}
\author{Harshit Joshi}
\affiliation{Univ. Bordeaux, CNRS, LOMA, UMR 5798, 33405 Talence, France.}
\author{Yacine Amarouchene}
\affiliation{Univ. Bordeaux, CNRS, LOMA, UMR 5798, 33405 Talence, France.}
\author{Thomas Salez}
\thanks{thomas.salez@cnrs.fr}
\affiliation{Univ. Bordeaux, CNRS, LOMA, UMR 5798, 33405 Talence, France.}
\affiliation{Mechanics Department, École Polytechnique, Institut Polytechnique de Paris, 91128 Palaiseau, France.}
\begin{abstract}
Fluid inertia can remain important in confined flows, even at low Reynolds number when convective inertia is negligible; yet its coupling with interface deformability is an open problem. Here, we theoretically investigate the high-frequency oscillations of an immersed sphere near an elastic wall in the lubrication regime. We first reconcile previous predictions for the added mass near a rigid wall by consistently matching the inner lubrication flow with the outer flow. Using this rigid-wall problem as a reference, we then develop a perturbative analysis of wall deformation and derive the leading elasto-inertial correction to the hydrodynamic force. 
Our results reveal that the coupling between interface softness and fluid inertia produces a dominant non-dissipative force contribution scaling as the power four of the frequency in the high-frequency limit, thus corresponding to the fourth time derivative of the sphere's position. This finding may have implications regarding the short-time dynamics of colloidal particles near deformable interfaces in biological and soft-matter settings.
\end{abstract}
\maketitle

The motion of objects in confined fluid environments is central to a wide range of biological and engineered systems.
When the confining interface is deformable, hydrodynamic stresses modify the gap geometry, which in turn alters the flow and the forces acting on the moving object. This feedback, known as elastohydrodynamic (EHD) coupling, makes object-interface interactions fundamentally different from their rigid counterparts~\cite{Karan2018,Bureau2023,rallabandi2024fluid}.
Understanding such a coupling is important across a range of settings, from cell transport through blood vessels~\cite{goldsmith1971red,grandchamp2013lift,Davies2018}, and lubrication in cartilaginous joints~\cite{grodzinsky1978electromechanical,mow1984fluid,mow2002mechano}, to the motion of colloidal particles near deformable substrates~\cite{Vakarelski2010},  contactless nanorheology~\cite{leroy2012hydrodynamic,Villey2013,kaveh2014hydrodynamic,wang2015out,guan2017noncontact,Vialar2019,Zhang2020}, and macroscopic friction~\cite{Saintyves2016,Rallabandi2018}

Fluid inertia is commonly neglected in lubrication and EHD theories. However, although convective inertia vanishes in the low-Reynolds-number limit, unsteady effects can remain important when the time scale of particle motion becomes comparable to the fluid vorticity diffusion time. The resulting hydrodynamic response then contains an added-mass contribution associated with fluid acceleration and a history force arising from the diffusion of vorticity~\cite{boussinesq1885resistance,basset1888treatise,franosch2011resonances,Fouxon2018a,Sorkin2026}. Their possible interplay with interface deformation raises a fundamental question: \textit{how does boundary softness modify the unsteady hydrodynamic response of a neighbouring particle?}

Answering the latter question represents a particular challenge. While viscous stresses are concentrated in the narrow gap between the particle and the wall, unsteady effects involve momentum transport and vorticity diffusion in the outer surrounding fluid. Consequently, a local lubrication description alone may not capture the complete inertial force, and a consistent theory must reconcile the inner flow with the outer flow. This difficulty is already apparent for a sphere oscillating near a rigid wall. High-frequency inertial effects have been studied using a matched-asymptotics approach by Chadwick \textit{et al.} \cite{chadwick2008high}, a potential-flow approach by Yang \cite{yang2010formula}, and a lubrication approach by Bigan \textit{et al.} \cite{bigan2024long}. These studies show that confinement can substantially modify the added mass relative to its bulk value, but the predictions of \cite{chadwick2008high} and \cite{yang2010formula} are inconsistent with each other, while \cite{bigan2024long} must invoke an ad-hoc cut-off bound in a diverging integral in order to circumvent the matching procedure.

Deformability introduces an additional difficulty into this problem. Already in the absence of fluid inertia, soft-lubrication theory~\cite{lee1982motion,berdan1982motion,sekimoto1993mechanism,Beaucourt2004,skotheim2004soft,urzay2007elastohydrodynamic,trouilloud2008soft,leroy2011hydrodynamic,Snoeijer2013,Pandey2016,Essink2021,bertin2022soft,jha2024capillary} has revealed a rich range of phenomena, including non-inertial lift, Magnus-like forces and dynamic adhesion, among other exotic features. Notably, these studies show that interface deformation can generate an added-mass-like contribution to the hydrodynamic response even when fluid inertia is neglected. Such effects have also been incorporated into descriptions of Brownian motion near soft interfaces \cite{ye2025brownian}. Besides, recent experiments have reported substantial soft interface contributions to Brownian dynamics \cite{fares2024observation}. However, the coupling between interface deformation and fluid inertia remains so far unresolved.

In this Letter, we investigate the unsteady hydrodynamic response of a sphere oscillating in a fluid near a deformable interface, in the high-frequency regime but at vanishing Reynolds number. Using the rigid-wall problem as a reference, we first reconcile the apparently discrepant predictions for the added mass in \cite{chadwick2008high} and \cite{yang2010formula}. We then develop a perturbation expansion based on the interface deformation rate, which governs the leading-order soft response in the high-frequency regime, and match the inner and outer flow solutions to derive the leading elasto-inertial correction to the hydrodynamic force. We find that the coupling between interface softness and fluid inertia produces a dominant non-dissipative contribution scaling as \(\omega^4\), where $\omega$ is the oscillation angular frequency, that is distinct from the added-mass response of the rigid-interface problem. 

\begin{figure}[t]
    \centering
    \includegraphics[
        width=\linewidth
    ]{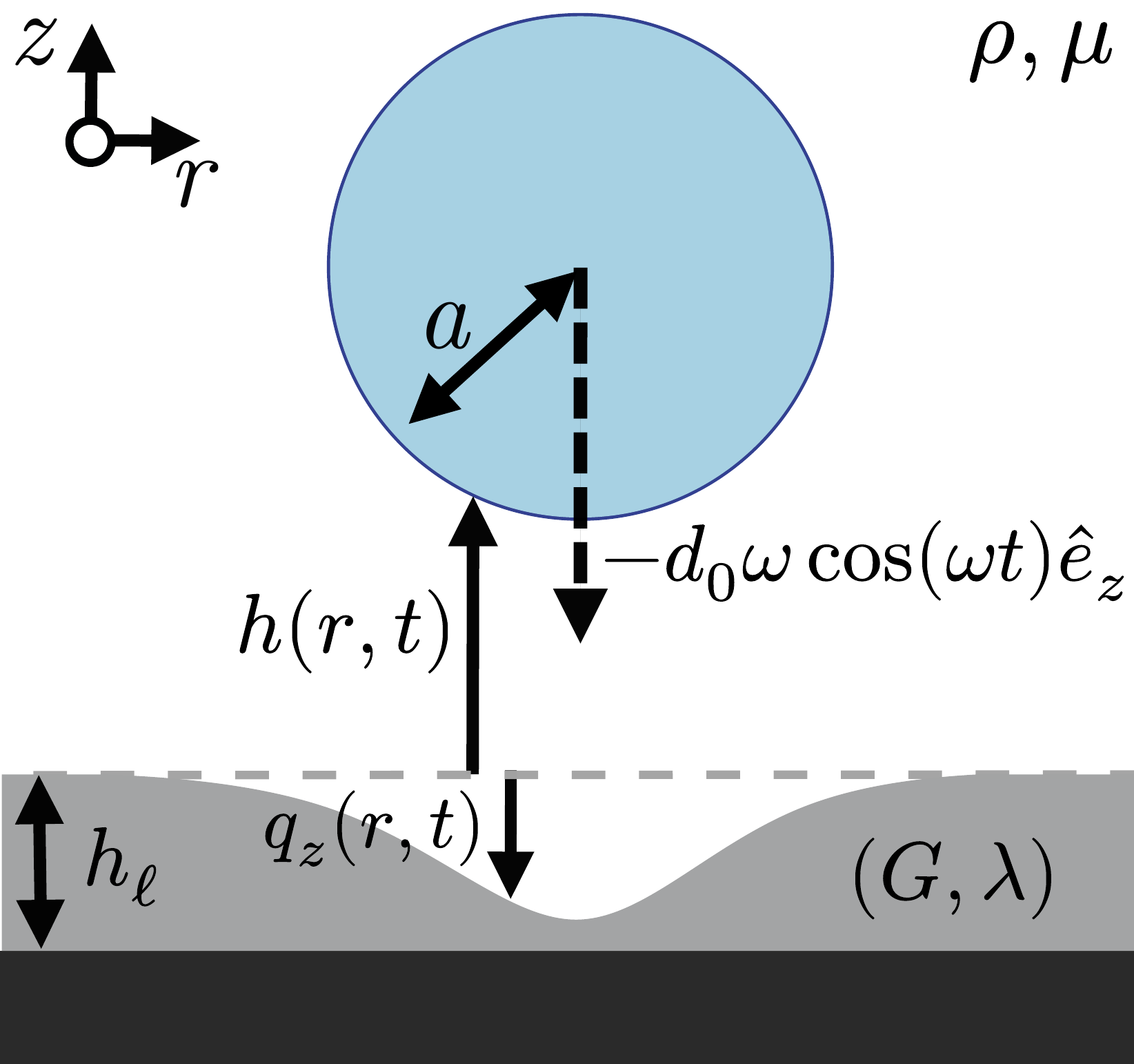}
    \caption{Schematic of the system. A sphere of radius $a$ oscillates with velocity $-d_0\omega cos(\omega t)\mathbf{e_z}$ in a fluid of viscosity $\mu$ and density $\rho$ with angular frequency $\omega$ and amplitude $d_0$, near a deformable wall of thickness $h_{\ell}$ and Lam\'e coefficients $G$ and $\lambda$. We denote by $h(r,t)$ the undeformed gap profile,  and by $q_z(r,t)$ the vertical deformation field of the wall. We introduce the time variable $t$, as well as cylindrical coordinates $(r,z)$, with the origin located on the undeformed surface indicated by the horizontal dashed line.}
    \label{fig:scheme_problem}
\end{figure}

We consider the system depicted in Fig.~\ref{fig:scheme_problem}. The fluid velocity and pressure fields are denoted by $\bm{u}=(u_r,u_z)$ and $p$, respectively. 
We non-dimensionalize the variables, as:
$R=r/a$,
$Z=z/a$,
$T=\omega t$,
$H=h/a$,
$P=p\, a/(\mu d_0\omega)$,
$\bm{U}=(U_R,U_Z)=\bm{u}/(d_0\omega)$,
 $Q_Z=q_z/a$, and $\delta=d_0/a$.
The time dependence of a variable $\mathcal{A}(\chi,T)$ is decomposed into Fourier modes as
$\mathcal{A}(\chi,T)=\Re\left[\sum_{n\geq1}\mathcal{A}^{(n)}(\chi)e^{-inT}\right]$,
where $\chi$ is any set of relevant coordinates. 

The dimensionless unsteady Stokes equations for an incompressible fluid, and a given Fourier mode, are:
\begin{subequations}
\label{conservation_equation}
    \begin{gather}
\nabla P^{(n)}
=
i\beta_n^2\bm{U}^{(n)}
+
\nabla^2\bm{U}^{(n)}\ ,
\\
\nabla\cdot\bm{U}^{(n)}
=
0\ ,
\end{gather}
\end{subequations}
where $\beta_n^2=n\beta^2=n(\rho\omega a^2)/\mu$.
The axisymmetric nature of the problem allows for a description in terms of the stream function $\Psi^{(n)}$, through
$U_R^{(n)}=(1/R)\,\partial_Z\Psi^{(n)}$
and
$U_Z=-(1/R)\,\partial_R\Psi^{(n)}$, as:
\begin{equation}
E^2(E^2+i\beta_n^2)\Psi^{(n)}=0\ ,
\label{conservation_equation_psi}
\end{equation}
with $E^2=\partial_R^2+\partial_Z^2-(1/R)\,\partial_R$.
The boundary conditions in the lab reference frame are given by:
\begin{equation}
\begin{aligned}
U_R\big|_{\mathrm{wall}} &= 0\ ,
&
U_R\big|_{\mathrm{sphere}} &= 0\ ,
\\
U_Z\big|_{\mathrm{wall}} &= \frac{1}{\delta}\partial_T Q_Z\ ,
&
U_Z\big|_{\mathrm{sphere}} &= -\cos (T)\ .
\end{aligned}
\label{general_BC}
\end{equation}

We assume $Q_Z\ll \epsilon$ and $\delta\ll\epsilon$, where $\epsilon=h_0/a$ is the small lubrication parameter with $h_0=h(0,0)$ the typical vertical fluid gap distance. We further assume that $Q_Z\ll\delta$. In this limit, one can perturb any physical quantity $\mathcal{A}$ as $\mathcal{A}\simeq\mathcal{A}_0+\mathcal{A}_1$, where $\mathcal{A}_1$ corresponds to the first-order correction in softness of the rigid solution $\mathcal{A}_0$.
For simplicity, we also assume that $Q_Z/\delta\gg\max(\delta/\epsilon,\epsilon)$ so that the only dominant soft correction is given by the boundary condition on $U_Z|_{\mathrm{wall}}$ (see details in the Supplementary Information (SI)). All together, the scale separation we impose is:
$\max(\delta^2/\epsilon,\delta\epsilon)\ll Q_Z\ll\delta\ll\epsilon\ll1$.

The height of the surface of the sphere, with respect to $Z=0$, is given by:
\begin{equation}
H=
1+\epsilon-\sqrt{1-R^2}
+
\mathcal{O}(\delta)\ ,
\label{sphere_height}
\end{equation}
for $R\leq1$. Besides, we assume a Winkler-like elastic response of the wall: \begin{equation}
Q_Z\simeq-\kappa P_0(Z=0)+\mathcal{O}(\kappa^2)\ ,
\label{wink}
\end{equation}
where we invoked the dimensionless elastic compliance:
\begin{equation}
\kappa
=
\frac{h_\ell}{2G+\lambda}
\frac{\mu d_0\omega}{a^2}\ .
\end{equation}
Such a description is valid when the thickness $h_{\ell}$ of the elastic material  is much smaller than the hydrodynamic radius $\sqrt{2h_0a}$, and when $ (h_\ell/\sqrt{h_0 a})^2\ll1-2\nu$,
where $\nu=\lambda/[2(\lambda+G)]$ is the Poisson ratio, so that compression dominates the shear-mediated contribution to the normal displacement~\cite{winkler1867theory, chandler2020validity}.
Therefore, to find the soft correction to the rigid-wall problem, one first needs to know the pressure field near the rigid wall.

We solve the rigid problem following the approach of Cooley and O'Neill \cite{cooley1969slow}, by constructing inner and outer solutions, which are respectively the solutions of the flow in the thin gap under the sphere and in the region far from the wall.
In the inner region, we define $\mathcal{O}(1)$ variables, as:
$\overline{Z}=Z/\epsilon$,
$\overline{H}=H/\epsilon$,
$\overline{R}=R/\sqrt{\epsilon}$,
$\overline{P}=\epsilon^2P$,
$\overline{U}_Z=U_Z$,
$\overline{U}_R=\sqrt{\epsilon}U_R$,
and
$\overline{\Psi}=\Psi/\epsilon$. Using these scalings and Eq.~\eqref{sphere_height}, we get
$\overline{H} \simeq 1+\overline{R}^2/2 + \mathcal{O}(\epsilon)+\mathcal{O}(\delta/\epsilon)$.
Non-negligible inertia in this region requires oscillations with high-enough frequency. We thus impose $\beta^2=\mathcal{O}(\epsilon^{-2})$, and introduce
$\alpha_n^2=n\alpha^2=\epsilon^2\beta_n^2$,
with $\alpha^2=\mathcal{O}(1)$.
Since $E^2\simeq\epsilon^{-2} \partial_{\overline{Z}}^2 + \mathcal{O}(\epsilon^{-1})$, Eqs.~\eqref{conservation_equation_psi} and  \eqref{general_BC} can be solved analytically.
The detailed expression of the $n$-th Fourier component $\overline{\Psi}_0^{(n)}$ of the inner stream function is given in the SI. 

\begin{figure}[t]
    \centering
    \includegraphics[
        width=\linewidth
    ]{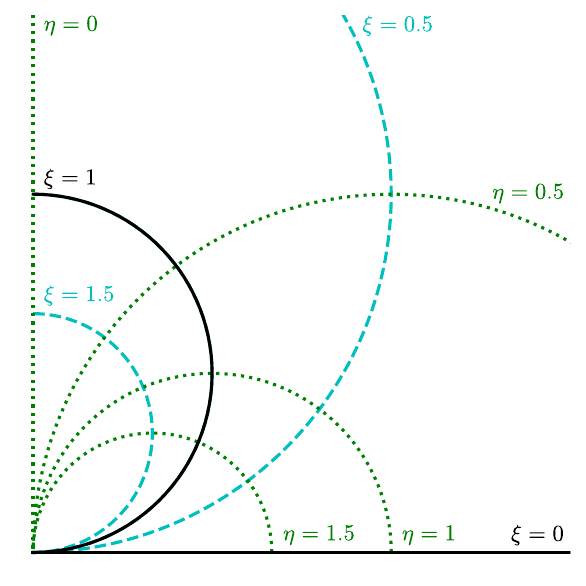}
    \caption{Representation of the tangent-sphere coordinates~\cite{oneill1967slow}.}
    \label{fig:tangential_coordinates}
\end{figure}

In the outer region, we use the tangent-sphere coordinates $(\xi,\eta)$, defined through~\cite{oneill1967slow} :
$R=2\eta/(\xi^2+\eta^2)$
and
$Z=2\xi/(\xi^2+\eta^2)$
(see Fig.~\ref{fig:tangential_coordinates}).
In this coordinate system, the undeformed wall corresponds to $\xi=0$, and the sphere surface corresponds to $\xi=1+\epsilon (\eta^2-1)/2 + \mathcal{O}(\epsilon^2)$. 
The outer solution then follows from the method proposed by Chadwick and Liao \cite{chadwick2008high}. The latter consists in first constructing a boundary-layer solution near the sphere and matching it with the potential-flow solution far from the sphere. However, doing so, we found a different result than \cite{chadwick2008high}, as described in the SI.

We further construct a boundary-layer solution for the outer stream function near the undeformed wall and match it with the potential-flow solution far from the wall. Indeed, this step is needed to obtain the pressure field in the outer region on the undeformed wall, which in turn gives the wall deformation through Eq.~\eqref{wink}. The detailed wall boundary-layer solution is provided in the SI. Equation~\eqref{conservation_equation} relates the pressure gradient to the stream function, in the Fourier representation, as:
\begin{equation}
\begin{pmatrix}
\partial_R P_0^{(n)}
\\
\partial_Z P_0^{(n)}
\end{pmatrix}
=
\frac{1}{R}
\begin{pmatrix}
\partial_Z E^2\Psi_0^{(n)}
+i\beta_n^2\partial_Z\Psi_0^{(n)}
\\
-\partial_R E^2\Psi_0^{(n)}
-i\beta_n^2\partial_R\Psi_0^{(n)}
\end{pmatrix}.
\label{eq_pressure_stream}
\end{equation}
Since the pressure vanishes far from the sphere, using the tangent-sphere coordinates with the wall boundary-layer solution and integrating the above equation gives the leading-order pressure field at the rigid wall in the outer region, as:
\begin{equation}
\widetilde{P}_0^{(n)}(\xi=0,\eta)
=
-\delta_{n,1}
\frac{i\alpha^2}{2}
\int_0^\eta \mathrm{d}s\, s b_{\mathrm{w},0}(\eta)\ ,
\label{pOuterEqn}
\end{equation}
where
$\delta_{i,j}$ is the Kronecker symbol, $b_{\mathrm{w},0}(\eta)=\int_0^\infty\mathrm{d}\lambda\,(2\lambda^2e^{-\lambda}/\sinh\lambda)\,J_1(\lambda\eta)$, and $\widetilde{P}_0^{(n)}=P_0^{(n)}\epsilon^2$. Hence, the pressure in the vicinity of the wall is on the order of $1/\epsilon^2$ relative to the bulk pressure, as in the inner region. This scaling results from the high-frequency limit, which gives rise to non-negligible inertial effects. We note that, as a consequence, when incorporating elasticity later on, the wall deformation will scale as $Q_Z = \mathcal{O}(\kappa/\epsilon^2)$.

Let us now evaluate the pressure field at the rigid wall in the inner solution. Using Eq.~\eqref{eq_pressure_stream} and the inner stream function obtained above, we get in the Fourier representation:
\begin{align}
\partial_{\overline{R}}\overline{P}_0^{(n)}
&=
\frac{
i\sqrt{-i}\,\alpha^3
}{
4
}
\,
\delta_{n,1}
\,
\frac{
\overline{R}
}{
(Y/2)-\tanh\!\left(
Y/2
\right)
}\ ,
\end{align}
where $Y= \sqrt{-i}\,\alpha\,(1+\overline{R}^2/2)$.
The pressure field at the wall in the inner region can thus be expressed, in Fourier representation, as:
\begin{align}
\overline{P}_0^{(n)}
(\overline{R},\overline{Z}=0)
&=
-\int_{\overline{R}}^{\overline{R}_0}
\textrm{d} \overline{R'}\,
\partial_{\overline{R'}}
\overline{P}_0^{(n)}
(\overline{R'},\overline{Z}=0)
\nonumber\\
&\quad
+\widetilde{P}_0^{(n)}
(\xi=0,\eta=\eta_0)\ ,
\label{eq:pInner}
\end{align}
where $\overline{R}_0=2/(\sqrt{\epsilon}\eta_0)$, with $\eta_0$ chosen in the range
$1\ll\eta_0\ll\epsilon^{-1/2}$
where the inner and outer stream functions match \cite{chadwick2008high}.
The constant of integration for the inner pressure ensures that the inner and outer pressure fields match in the matching region.

We now turn to the calculation of the hydrodynamic force acting on the sphere in the rigid case.
The vertical hydrodynamic force $f^{(n)}$, first derived by Stimson and Jeffery \cite{stimson1926motion}, and extended later by Chadwick and Liao \cite{chadwick2008high}, following the method of Happel and Brenner \cite{happel2012low}, can be expressed as follows \footnote{Note the correction in the sign of the second term from \cite{chadwick2008high}}:
\begin{align}
f^{(n)}
&=
\int_\gamma\mathrm{d}\gamma\,
R^3
\partial_{\textrm{n}}
\left(
\frac{1}{R^2}E^2\Psi^{(n)}
\right)\ ,
\nonumber\\
&\quad
+i\beta_n^2
\int_\gamma
\mathrm{d}\gamma\, R\,
\partial_{\textrm{n}}\Psi^{(n)}\ ,
\label{force}
\end{align}
where $\gamma$ is a path that goes from the top to the bottom poles of the sphere along the meridian in the $rz$ plane, and $\partial_\textrm{n}$ represents the derivative with respect to the outward unit vector normal to the sphere.
The second integral in Eq.~\eqref{force} can be determined exactly using the flow boundary conditions on the sphere, giving $\frac{4}{3} i \beta^2_1\delta_{n,1}$.
Separating the first integral into an inner integral going from $\overline{R}=0$ to $\overline{R}_0$ and an outer integral going from $\eta=\eta_0$ to $0$, and using the asymptotic forms of the stream functions in the inner and outer regions, leads to the following result for the force at dominant order in $\epsilon$, in the rigid case and in the Fourier representation:
\begin{equation}
f_0^{(n)}
=
\delta_{n,1}
\frac{2i\alpha^2}{\epsilon^2}
\left[
\frac{2}{3}-\zeta(3)
\right] + \mathcal{O}(1/\epsilon)\ ,
\label{rigid_force}
\end{equation}
where $\zeta$ denotes the Riemann zeta function, and where the subscript $0$ indicates the rigid case.
The Kronecker delta implies that the dominant flow response for the rigid-wall problem has one dominant mode: the same one as the oscillating sphere.
The obtained force can be interpreted as an added-mass force, with an added mass expressed in real units as
$M_{\textrm{f}}\,[3\zeta(3)/2-1]$,
where $M_{\textrm{f}}$ is the fluid mass displaced by the motion of the sphere. This result reconciles the results of Chadwick \cite{chadwick2008high} and Yang \cite{yang2010formula}, which have been the subject of recent experimental \cite{ferreira2026quantitative} and numerical \cite{thomas2026broadband} studies.

Let us now turn to the soft case. The softness-induced correction $\Psi_1^{(n)}$ to the stream function, in Fourier representation, satisfies Eq.~\eqref{conservation_equation_psi}. The only non-zero boundary conditions are given by:
\begin{equation}
\Psi_{\mathrm{i},1}^{(n)}(\overline{R},\overline{Z}=0)
=
-\frac{in\kappa}{\delta\epsilon^2}
\int_0^{\overline{R}}\mathrm{d}R'\,
R'\,\overline{P}_0^{(n)}(R',\overline{Z}=0)\ ,
\end{equation}
for the inner (``i") region, and by:
\begin{equation}
\Psi_{\mathrm{o},1}^{(n)}(\xi=0,\eta)
=
-\frac{4in\kappa}{\delta\epsilon^2}
\int_\eta^\infty\mathrm{d}s\,
\frac{\mathcal{P}_0^{(n)}(\xi=0,\eta=s)}{s^3},
\end{equation}
for the outer (``o") region, where $\mathcal{P}$ is defined as the general pressure field that is equal to $\widetilde{P}$ in the outer region and $\overline{P}$ in the inner region. By construction, the function $\mathcal{P}$ is continuous.
The solutions $\Psi_{\mathrm{o},1}^{(n)}$ and $\Psi_{\mathrm{i},1}^{(n)}$ are obtained like in the rigid case (see SI).
In the matching region, and in the limit close to the sphere, $\overline{Z},\overline{R}\gg1$ and $|1+\overline{R}^2/2-\overline{Z}|\ll1$, it can be shown that $\Psi_{\mathrm{i},1}^{(n)}\simeq\Psi_{\mathrm{o},1}^{(n)}$.
Combining these stream functions with Eq.~\eqref{force}, after some algebra (see SI), leads to:
\begin{equation}
f_{\mathrm{i},1}^{(n)}
\underset{\overline{R}_0\gg1}{\simeq}
-\delta_{n,1}
\frac{2i\kappa\alpha^4}{\delta\epsilon^4}
\frac{\ln(\eta_0)}{\eta_0^2}\ ,
\label{inner_force}
\end{equation}
and:
\begin{align}
f_{\mathrm{o},1}^{(n)}
&
\underset{\eta_0\gg1}{\simeq}
-\delta_{n,1}
\zeta(4)
\frac{i\kappa\alpha^4}{\delta\epsilon^4}
+
\delta_{n,1}
\frac{2i\kappa\alpha^4}{\delta\epsilon^4}
\frac{\ln(\eta_0)}{\eta_0^2}\ .
\label{outer_force}
\end{align}

Finally, combining Eqs.~\eqref{rigid_force}, \eqref{inner_force} and \eqref{outer_force} yields the dominant hydrodynamic force:
\begin{equation}
f^{(n)}
\simeq
-\frac{i\alpha^2}{\epsilon^2}
\delta_{n,1}
\left[
2\zeta(3)
-\frac{4}{3}
+
\zeta(4)
\frac{\kappa}{\delta\epsilon^2}
\alpha^2
\right] + \mathcal{O}(1/\epsilon)\ .
\label{nondim_force}
\end{equation}
The softness-induced correction scales as $\mathcal{O}(\kappa/(\delta\epsilon^4))$, which is small compared to the dominant force scaling as $\mathcal{O}(1/\epsilon^2)$, consistent with the perturbative expansion and the scale separation.
Putting back dimensions, one has the following expression of the dominant vertical hydrodynamic force in the high-frequency limit:
\begin{align}
F
&\simeq
-\sin(\omega t)M_{\textrm{f}} d_0\omega^2
\left[
\frac{3}{2}\zeta(3)-1
+
\frac{3}{4}\zeta(4)
\frac{h_{\ell}}{2G+\lambda}
\rho a\omega^2
\right].
\label{dim_force}
\end{align}
Interestingly, the softness-induced correction corresponds to a non-dissipative inertial response, like the classical added-mass term. However, it behaves as $\sim d_0\omega^4$, which formally corresponds to a fourth time derivative of the particle position instead of the usual second derivative, hence differing from a strict added-mass effect. Furthermore, we stress that, as in the rigid case, the dominant response exhibits no mode coupling. 
However, if we instead adopted the scaling $\kappa/{(\delta\epsilon^2)}\sim\max(\delta/\epsilon, \epsilon)$, the geometric corrections of order $\mathcal{O}(\delta/\epsilon)$ may be retained in the boundary conditions. These corrections involve products of the sphere's oscillatory displacement with spatial derivatives of the fluid velocity, generating a mean component and second harmonics, as well as phase shifts in the resulting oscillatory response. However, even in this case, the leading-order first-harmonic contribution remains unchanged as compared to Eq.~\eqref{nondim_force}.

We investigated the effect of fluid inertia on a sphere oscillating near an elastic wall in the high-frequency lubrication regime. First, for a rigid wall, we showed that the analysis of Chadwick and Liao \cite{chadwick2008high} requires a correction at leading order. Including this correction yielded a fluid added-mass effect consistent with the potential-flow result of Yang \cite{yang2010formula}, hence solving the literature controversy on that matter. Second, using such a rigid-wall result as a reference, we performed a perturbative analysis in the dimensionless compliance, in order to determine the leading effect of wall softness on the  hydrodynamic response. We found that the coupling between interface softness and fluid inertia generates a non-dissipative force contribution proportional to the fourth time derivative of the sphere's position. Our finding hence complements the added-mass effects arising in : i) unsteady hydrodynamics near rigid boundaries~\cite{chadwick2008high,yang2010formula,ferreira2026quantitative}; and ii) soft lubrication without fluid inertia \cite{salez2015elastohydrodynamics,bertin2022soft,jha2024capillary,ye2025brownian}. The elasto-inertial force we uncovered may be accessible experimentally using atomic force microscopy~\cite{zhang2023unsteady,Palacios-Alonso2025,ferreira2026quantitative}. More broadly, it could influence the short-time dynamics and statistical properties of Brownian particles in complex soft-matter and biological environnements.

\begin{acknowledgments}
The authors thank Quentin Ferreira, Anirban Chatterjee, and Aditya Jha for interesting discussions. They acknowledge financial support from the European Union through the European Research Council under EMetBrown (ERC-CoG-101039103) grant. They also acknowledge financial support from the University of Bordeaux under BioFiX Prematuration grant. Besides, they acknowledge the support from the R\'eseaux de Recherche Impulsion Frontiers of Life which received financial support from the French government in the framework of the University of Bordeaux's France 2030 program. Finally, they thank the Soft Matter Collaborative Research Unit, Frontier Research Center for Advanced Material and Life Science, Faculty of Advanced Life Science at Hokkaido University, Sapporo, Japan, and the CNRS International Research Network between France and India on Hydrodynamics at small scales: From soft matter to bioengineering.
\end{acknowledgments}
\bibliographystyle{unsrt}
\bibliography{Batalha2026}

\begin{thebibliography}{10}

\bibitem{Karan2018}
P.~Karan, J.~Chakraborty, and S.~Chakraborty.
\newblock Small-scale flow with deformable boundaries.
\newblock {\em J. Indian Inst. Sci.}, 98:159, 2018.

\bibitem{Bureau2023}
Lionel Bureau, Gwennou Coupier, and Thomas Salez.
\newblock Lift at low reynolds number.
\newblock {\em The European Physical Journal E}, 46(11):111, 2023.

\bibitem{rallabandi2024fluid}
Bhargav Rallabandi.
\newblock Fluid-elastic interactions near contact at low reynolds number.
\newblock {\em Annual Review of Fluid Mechanics}, 56:491--519, 2024.

\bibitem{goldsmith1971red}
HL~Goldsmith.
\newblock Red cell motions and wall interactions in tube flow.
\newblock In {\em Federation proceedings}, volume~30, pages 1578--1590, 1971.

\bibitem{grandchamp2013lift}
Xavier Grandchamp, Gwennou Coupier, Aparna Srivastav, Christophe Minetti, and
  Thomas Podgorski.
\newblock Lift and down-gradient shear-induced diffusion in red blood cell
  suspensions.
\newblock {\em Physical review letters}, 110(10):108101, 2013.

\bibitem{Davies2018}
Heather~S. Davies, Delphine Debarre, Nouha El~Amri, Claude Verdier, Ralf~P.
  Richter, and Lionel Bureau.
\newblock Elastohydrodynamic lift at a soft wall.
\newblock {\em Phys. Rev. Lett.}, 120:198001, 2018.

\bibitem{grodzinsky1978electromechanical}
AJ~Grodzinsky, H~Lipshitz, and MJ~Glimcher.
\newblock Electromechanical properties of articular cartilage during
  compression and stress relaxation.
\newblock {\em Nature}, 275(5679):448--450, 1978.

\bibitem{mow1984fluid}
Van~C Mow, Mark~H Holmes, and W~Michael Lai.
\newblock Fluid transport and mechanical properties of articular cartilage: a
  review.
\newblock {\em Journal of biomechanics}, 17(5):377--394, 1984.

\bibitem{mow2002mechano}
Van~C Mow and X~Edward Guo.
\newblock Mechano-electrochemical properties of articular cartilage: their
  inhomogeneities and anisotropies.
\newblock {\em Annual review of biomedical engineering}, 4(1):175--209, 2002.

\bibitem{Vakarelski2010}
Ivan~U. Vakarelski, Rogerio Manica, Xiaosong Tang, Sean~J. O'Shea, Geoffrey~W.
  Stevens, Franz Grieser, Raymond~R. Dagastine, and Derek Y.~C. Chan.
\newblock Dynamic interactions between microbubbles in water.
\newblock {\em Proc. Nat. Acad. Sci.}, 107(25):11177--11182, 2010.

\bibitem{leroy2012hydrodynamic}
Samuel Leroy, Audrey Steinberger, C{\'e}cile Cottin-Bizonne, Fr{\'e}d{\'e}ric
  Restagno, Liliane L{\'e}ger, and Elisabeth Charlaix.
\newblock Hydrodynamic interaction between a spherical particle and an elastic
  surface: a gentle probe for soft thin films.
\newblock {\em Physical review letters}, 108(26):264501, 2012.

\bibitem{Villey2013}
Richard Villey, Emmanuelle Martinot, C\'ecile Cottin-Bizonne, Magali
  Phaner-Goutorbe, Liliane L\'eger, Fr\'ed\'eric Restagno, and Elisabeth
  Charlaix.
\newblock Effect of surface elasticity on the rheology of nanometric liquids.
\newblock {\em Phys. Rev. Lett.}, 111:215701, 2013.

\bibitem{kaveh2014hydrodynamic}
Farzaneh Kaveh, Javed Ally, Michael Kappl, and Hans-J\"urgen Butt.
\newblock Hydrodynamic force between a sphere and a soft, elastic surface.
\newblock {\em Langmuir}, 30(39):11619--11624, 2014.

\bibitem{wang2015out}
Yumo Wang, Charles Dhong, and Joelle Frechette.
\newblock Out-of-contact elastohydrodynamic deformation due to lubrication
  forces.
\newblock {\em Physical review letters}, 115(24):248302, 2015.

\bibitem{guan2017noncontact}
Dongshi Guan, Chlo{\'e} Barraud, Elisabeth Charlaix, and Penger Tong.
\newblock Noncontact viscoelastic measurement of polymer thin films in a liquid
  medium using long-needle atomic force microscopy.
\newblock {\em Langmuir}, 33(6):1385--1390, 2017.

\bibitem{Vialar2019}
Pierre Vialar, Pascal Merzeau, Suzanne Giasson, and Carlos Drummond.
\newblock Compliant surfaces under shear: Elastohydrodynamic lift force.
\newblock {\em Langmuir}, 35(48):15605--15613, 2019.

\bibitem{Zhang2020}
Zaicheng Zhang, Vincent Bertin, Muhammad Arshad, Elie Rapha{\"e}l, Thomas
  Salez, and Abdelhamid Maali.
\newblock Direct measurement of the elastohydrodynamic lift force at the
  nanoscale.
\newblock {\em Physical review letters}, 124(5):054502, 2020.

\bibitem{Saintyves2016}
Baudouin Saintyves, Theo Jules, Thomas Salez, and L.~Mahadevan.
\newblock Self-sustained lift and low friction via soft lubrication.
\newblock {\em Proc. Nat. Acad. Sci.}, 113(21):5847--5849, 2016.

\bibitem{Rallabandi2018}
B.~Rallabandi, N.~Oppenheimer, M.~Y.~B. Zion, and H.~A. Stone.
\newblock Membrane-induced hydroelastic migration of a particle surfing its own
  wave.
\newblock {\em Nature Phys.}, 14:1211, 2018.

\bibitem{boussinesq1885resistance}
Joseph Boussinesq.
\newblock Sur la resistance qu'oppose un fluide indefini en repos, sans
  pesanteur, au mouvement varie d'une sphere solide qu'il mouille sur toute sa
  surface, quand les vitesses restent bien continues et assez faibles pour que
  leurs carres et produits soient negligiables.
\newblock {\em CR Acad. Sc. Paris}, 100:935--937, 1885.

\bibitem{basset1888treatise}
Alfred~Barnard Basset.
\newblock {\em A treatise on hydrodynamics: with numerous examples}, volume~2.
\newblock Deighton, Bell and Company, 1888.

\bibitem{franosch2011resonances}
Thomas Franosch, Matthias Grimm, Maxim Belushkin, Flavio~M Mor, Giuseppe Foffi,
  L{\'a}szl{\'o} Forr{\'o}, and Sylvia Jeney.
\newblock Resonances arising from hydrodynamic memory in brownian motion.
\newblock {\em Nature}, 478(7367):85--88, 2011.

\bibitem{Fouxon2018a}
Itzhak Fouxon and Alexander Leshansky.
\newblock Fundamental solution of unsteady {{Stokes}} equations and force on an
  oscillating sphere near a wall.
\newblock {\em Physical Review E}, 98(6):063108, December 2018.

\bibitem{Sorkin2026}
Benjamin Sorkin, G{\"u}nther Turk, and Howard~A Stone.
\newblock Brownian motion with geometry-dependent hydrodynamic memory.
\newblock {\em arXiv preprint arXiv:2609.17678}, 2026.

\bibitem{chadwick2008high}
Richard~S Chadwick and Zhijie Liao.
\newblock High-frequency oscillations of a sphere in a viscous fluid near a
  rigid plane.
\newblock {\em SIAM review}, 50(2):313--322, 2008.

\bibitem{yang2010formula}
F-L Yang.
\newblock A formula for the wall-amplified added mass coefficient for a solid
  sphere in normal approach to a wall and its application for such motion at
  low reynolds number.
\newblock {\em Physics of Fluids}, 22(12), 2010.

\bibitem{bigan2024long}
Nathan Bigan, Mathieu Liz{\'e}e, Marc Pascual, Antoine Nigu{\`e}s, Lyd{\'e}ric
  Bocquet, and Alessandro Siria.
\newblock Long range signature of liquid's inertia in nanoscale drainage flows.
\newblock {\em Soft Matter}, 20(44):8804--8811, 2024.

\bibitem{lee1982motion}
SH~Lee and LG~Leal.
\newblock The motion of a sphere in the presence of a deformable interface: Ii.
  a numerical study of the translation of a sphere normal to an interface.
\newblock {\em Journal of Colloid and Interface Science}, 87(1):81--106, 1982.

\bibitem{berdan1982motion}
C.~Berdan~{II} and L.~G. Leal.
\newblock Motion of a sphere in the presence of a deformable interface: I.
  perturbation of the interface from flat: The effects on drag and torque.
\newblock {\em Journal of Colloid and Interface Science}, 87(1):62--80, 1982.

\bibitem{sekimoto1993mechanism}
K~Sekimoto and L~Leibler.
\newblock A mechanism for shear thickening of polymer-bearing surfaces:
  elasto-hydrodynamic coupling.
\newblock {\em Europhysics Letters}, 23(2):113, 1993.

\bibitem{Beaucourt2004}
J.~Beaucourt, T.~Biben, and C.~Misbah.
\newblock Optimal lift force on vesicles near a compressible substrate.
\newblock {\em Europhys. Lett.}, 67:676, 2004.

\bibitem{skotheim2004soft}
Jan~M Skotheim and Laksminarayanan Mahadevan.
\newblock Soft lubrication.
\newblock {\em Physical review letters}, 92(24):245509, 2004.

\bibitem{urzay2007elastohydrodynamic}
Javier Urzay, Stefan~G Llewellyn~Smith, and Beverley~J Glover.
\newblock The elastohydrodynamic force on a sphere near a soft wall.
\newblock {\em Physics of Fluids}, 19(10):103106, 2007.

\bibitem{trouilloud2008soft}
Renaud Trouilloud, S~Yu Tony, AE~Hosoi, and Eric Lauga.
\newblock Soft swimming: Exploiting deformable interfaces for low reynolds
  number locomotion.
\newblock {\em Physical review letters}, 101(4):048102, 2008.

\bibitem{leroy2011hydrodynamic}
Samuel Leroy and Elisabeth Charlaix.
\newblock Hydrodynamic interactions for the measurement of thin film elastic
  properties.
\newblock {\em Journal of Fluid Mechanics}, 674:389--407, 2011.

\bibitem{Snoeijer2013}
J.~H. Snoeijer, J.~Eggers, and C.~H. Venner.
\newblock Similarity theory of lubricated hertzian contacts.
\newblock {\em Phys. Fluids}, 25(10):101705, 2013.

\bibitem{Pandey2016}
Anupam Pandey, Stefan Karpitschka, Cornelis~H. Venner, and Jacco~H. Snoeijer.
\newblock Lubrication of soft viscoelastic solids.
\newblock {\em Journal of Fluid Mechanics}, 799:433?447, 2016.

\bibitem{Essink2021}
Martin~H Essink, Anupam Pandey, Stefan Karpitschka, Cornelis~H Venner, and
  Jacco~H Snoeijer.
\newblock Regimes of soft lubrication.
\newblock {\em J. Fluid Mech.}, 915, 2021.

\bibitem{bertin2022soft}
Vincent Bertin, Yacine Amarouchene, Elie Rapha{\"e}l, and Thomas Salez.
\newblock Soft-lubrication interactions between a rigid sphere and an elastic
  wall.
\newblock {\em Journal of fluid mechanics}, 933:A23, 2022.

\bibitem{jha2024capillary}
Aditya Jha, Yacine Amarouchene, and Thomas Salez.
\newblock Capillary lubrication of a spherical particle near a fluid interface.
\newblock {\em Journal of Fluid Mechanics}, 1001:A58, 2024.

\bibitem{ye2025brownian}
Yilin Ye, Yacine Amarouchene, Rapha{\"e}l Sarfati, David~S Dean, and Thomas
  Salez.
\newblock Brownian motion near a soft surface.
\newblock {\em Comptes Rendus. Physique}, 26(G1):619--630, 2025.

\bibitem{fares2024observation}
Nicolas Fares, Maxime Lavaud, Zaicheng Zhang, Aditya Jha, Yacine Amarouchene,
  and Thomas Salez.
\newblock Observation of brownian elastohydrodynamic forces acting on confined
  soft colloids.
\newblock {\em Proceedings of the National Academy of Sciences},
  121(42):e2411956121, 2024.

\bibitem{winkler1867theory}
Emil Winkler.
\newblock Theory of elasticity and strength.
\newblock {\em Dominicus Prague}, 36, 1867.

\bibitem{chandler2020validity}
Thomas~GJ Chandler and Dominic Vella.
\newblock Validity of winkler’s mattress model for thin elastomeric layers:
  beyond poisson’s ratio.
\newblock {\em Proceedings. Mathematical, Physical, and Engineering Sciences},
  476(2242):20200551, 2020.

\bibitem{cooley1969slow}
MDA Cooley and ME~O'{N}eill.
\newblock On the slow motion generated in a viscous fluid by the approach of a
  sphere to a plane wall or stationary sphere.
\newblock {\em Mathematika}, 16(1):37--49, 1969.

\bibitem{oneill1967slow}
ME~O'{N}eill and K~Stewartson.
\newblock On the slow motion of a sphere parallel to a nearby plane wall.
\newblock {\em Journal of Fluid Mechanics}, 27(4):705--724, 1967.

\bibitem{stimson1926motion}
Margaret Stimson and George~Barker Jeffery.
\newblock The motion of two spheres in a viscous fluid.
\newblock {\em Proceedings of the Royal Society of London. Series A, Containing
  Papers of a Mathematical and Physical Character}, 111(757):110--116, 1926.

\bibitem{happel2012low}
John Happel and Howard Brenner.
\newblock {\em Low Reynolds number hydrodynamics: with special applications to
  particulate media}.
\newblock Springer Science \& Business Media, 2012.

\bibitem{Note1}
Note the correction in the sign of the second term from \cite
  {chadwick2008high}.

\bibitem{ferreira2026quantitative}
Quentin Ferreira, Pablo Palacios-Alonso, Harshit Joshi, Rafael
  Delgado-Buscalioni, Yacine Amarouchene, and Thomas Salez.
\newblock Quantitative measurement of fluid inertial effects in confined
  brownian motion.
\newblock {\em arXiv preprint arXiv:2606.09193}, 2026.

\bibitem{thomas2026broadband}
Quentin Thomas, Clara~Mefo Sop, Maxime Lavaud, Yacine Amarouchene, Thomas
  Salez, and Pascal Damman.
\newblock Broadband molecular dynamics simulation of fluid inertial effects in
  confined brownian motion.
\newblock {\em arXiv preprint arXiv:2606.24291}, 2026.

\bibitem{salez2015elastohydrodynamics}
Thomas Salez and Lakshminarayanan Mahadevan.
\newblock Elastohydrodynamics of a sliding, spinning and sedimenting cylinder
  near a soft wall.
\newblock {\em Journal of Fluid Mechanics}, 779:181--196, 2015.

\bibitem{zhang2023unsteady}
Zaicheng Zhang, Vincent Bertin, Martin~H Essink, Hao Zhang, Nicolas Fares,
  Zaiyi Shen, Thomas Bickel, Thomas Salez, and Abdelhamid Maali.
\newblock Unsteady drag force on an immersed sphere oscillating near a wall.
\newblock {\em Journal of Fluid Mechanics}, 977:A21, 2023.

\bibitem{Palacios-Alonso2025}
Pablo {Palacios-Alonso}, Ra{\'u}l~P{\'e}rez Pel{\'a}ez, and Rafael
  {Delgado-Buscalioni}.
\newblock Fast spectral solver for viscoelastic structures under oscillatory
  flow in free space or wall-bounded domains: {{Applications}} to quartz
  crystal microbalance and force spectroscopy.
\newblock {\em The Journal of Chemical Physics}, 163(19):194109, November 2025.

\end{thebibliography}
\end{document}